\documentclass{PoS}
\usepackage{subcaption}

\usepackage{graphicx}
\usepackage{amsmath}
\usepackage{bm}
\usepackage{amssymb}
\usepackage{braket}
\usepackage{slashed}
\usepackage{enumitem}
\usepackage{bbm}

\title{Pion Form Factor with Overlap Fermion}

\ShortTitle{Pion Form Factor with Overlap Fermion}

\author{\speaker{Gen Wang}\\
        Dept. of Physics and Astronomy, University of Kentucky, Lexington, KY 40506, USA \\
        E-mail: \email{genwang27@uky.edu}}

\author{Jian Liang \\
        Dept. of Physics and Astronomy, University of Kentucky, Lexington, KY 40506, USA \\
        E-mail: \email{jian.liang@uky.edu}}

\author{Terrence Draper \\
        Dept. of Physics and Astronomy, University of Kentucky, Lexington, KY 40506, USA \\
        E-mail: \email{draper@pa.uky.edu}}

\author{Keh-Fei Liu \\
        Dept. of Physics and Astronomy, University of Kentucky, Lexington, KY 40506, USA \\
        E-mail: \email{liu@pa.uky.edu}}

\author{Yi-Bo Yang \\
        Institute of Theoretical Physics, Chinese Academy of Sciences, Beijing 100190, China \\
        E-mail: \email{ybyang@itp.ac.cn}

  \begin{center}
    \large{
      \vspace*{0.4cm}
      \includegraphics[scale=0.20]{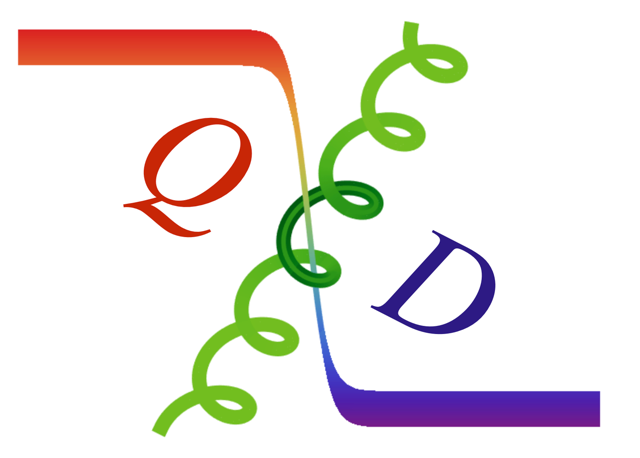}\\
      \vspace*{0.4cm}
      ($\chi$QCD Collaboration)
    }
  \end{center}
}
\abstract{
We present a calculation of the pion form factor using overlap fermions on 2+1-flavor domain-wall configurations on a $24^3\times 64$ lattice with $a=0.11 \, {\rm{fm}}$ and on a $32^3 \times 64$ lattice with $a=0.143 \, {\rm{fm}}$ generated by the RBC/UKQCD collaboration. Using the multi-mass algorithm, a simulation has been done with various valence quark masses with a range of space-like $Q^2$ from 0.0 to 0.6 ${\rm{GeV^2}}$.
}

\FullConference{The 36th Annual International Symposium on Lattice Field Theory - LATTICE2018\\
		22-28 July, 2018\\
		Michigan State University, East Lansing, Michigan, USA.}

\begin{document}

\section{Introduction} \label{intro}
    The charged pion electromagnetic form factor can be probed by a vector current of the photon. The experimental error of the pion electromagnetic form factor is around $1\%$ in the small $q^2$ region with mean squared radius $\langle r^2 \rangle =0.452(11) \ {\rm{fm^2}}$ as given by the Particle Data Group (PDG)~\cite{PhysRevD.98.030001}. Such a form factor can be calculated using Lattice QCD, but a precise measurement using Lattice QCD near the physical pion mass will be needed to control dominant systematic errors coming from its strong quark mass dependence. Decades of improving computational power along with the development of Lattice QCD theory and algorithms makes it possible to carry out a measurement from higher pion mass to the physical region~\cite{brommel_pion_2007-1,boyle_pions_2008-1,nguyen_electromagnetic_2011-1,jlqcd_collaboration_pion_2012,bali_nucleon_2013-1,brandt_pion_2013-1,jlqcd_collaboration_light_2016,koponen_size_2016} with various actions. It has been observed that in these calculations the values of $\langle r^2 \rangle$ at higher pion masses are significantly lower than the experimental value before chiral extrapolation~\cite{brandt_electromagnetic_2013}. On the other hand, one calculation using Highly Improved Staggered Quark formalism~\cite{koponen_size_2016} directly at the physical pion mass agrees with experiment.

    In order to study the above strong quark mass dependence of the pion form factor, the overlap fermion with multi-mass algorithm will be a good option compared to other formulations. By using existing grid source propagators with several quark masses generated by our $\chi{\rm{QCD}}$ collaboration using overlap fermions, a straightforward calculation of the vector current three-point functions and corresponding two-point functions has been carried out to obtain $\langle r^2 \rangle$ with less than $4\%$ error at several valence pion masses ranging from $147 \ {\rm{MeV}}$ to $389 \ {\rm{MeV}}$ on two lattices.

\section{Lattice Setup and Fitting Results}
    In this work, we use overlap fermions on 2+1-flavor domain-wall configurations on a $24^3\times 64$ lattice (24I) with Iwasaki gauge action and on a $32^3 \times 64$ lattice (32ID) with Iwasaki plus the Dislocation Suppressing Determinant Ratio (DSDR) gauge action~\cite{boyle_low_2016} listed in Table~\ref{tab:lattice}. For pion correlation functions, grid sources without low mode substitution (LMS) have been used in this production as suggested in reference~\cite{li_overlap_2010}. With a careful treatment of the backward pion propagation, we are able to use two time sources with separation of half of total number of time slices ($T/2$) to double the statistics. On 24I, a Gaussian-smeared source~\cite{degrand_wave-function_1991} has been applied with root mean square (RMS) radius $1.18 \ {\rm{fm}}$ along with the same smeared sink. On 32ID, a box-smeared source~\cite{allton_lattice_1991} has been applied with box half size $1.0 \ {\rm{fm}}$ along with the same smeared sink. Such a two-point function will be fitted using Eq.~(\ref{eq:2pt}).

\begin{eqnarray} \label{eq:2pt}
\begin{aligned}
C_{2pt}(t,\vec{p}) = & \sum_{\vec{x}} e^{-i {{\vec{p}}} \cdot {\vec{x}}} \braket{{\rm{T}}[\chi_{\pi^+}({\vec{x}},t) \chi_{\pi^+}^{\dagger}({\vec{0}},[0,T/2])} \\
\approx & \frac{m |Z_{\vec{p}}|^2}{E} (e^{-E t}+e^{-E (T/2-t)}) (1 + e^{-E (T/2)}) \\
& + \frac{m |Z^1_{\vec{p}}|^2}{E^1}  (e^{-E^1 t}+e^{-E^1 (T/2-t)}) (1 + e^{-E^1 (T/2)}) \\
\end{aligned}
\end{eqnarray}
$\chi_{\pi^+}(\vec{x},t) = \bar{u} (\vec{x},t) \gamma_{5} d (\vec{x},t)$ is the interpolating field of the pion, $E$ is the pion energy at momentum $\vec{p}$ and $Z_{\vec{p}}$ is the spectral weight for smeared source and sink at momentum $\vec{p}$ and $E^1$ and $Z^1_{\vec{p}}$ are the corresponding energy and spectral weight of the first excited state.

\begin{table}
  \centering{
  \begin{tabular}{| c | c | c | c | c | c | }
    \hline
    Lattice & Size & $a\ ({\rm{fm}})$ &  $L\ ({\rm{fm}})$ & Pion ({\rm{MeV}}) & $m_\pi L$ \\
    \hline
    24I     & $24^3\times 64$ & $0.11 $ & $2.64  $ & $337$ & $4.5$  \\
    \hline
    32ID    & $32^3\times 64$ & $0.143$ & $4.58 $ & $171$ & $3.97$ \\
    \hline
  \end{tabular}
  \caption{ Lattice parameters of 24I and 32ID}
  \label{tab:lattice}
  }
\end{table}

\begin{figure}
  \centering
  \includegraphics[width=0.45\linewidth]{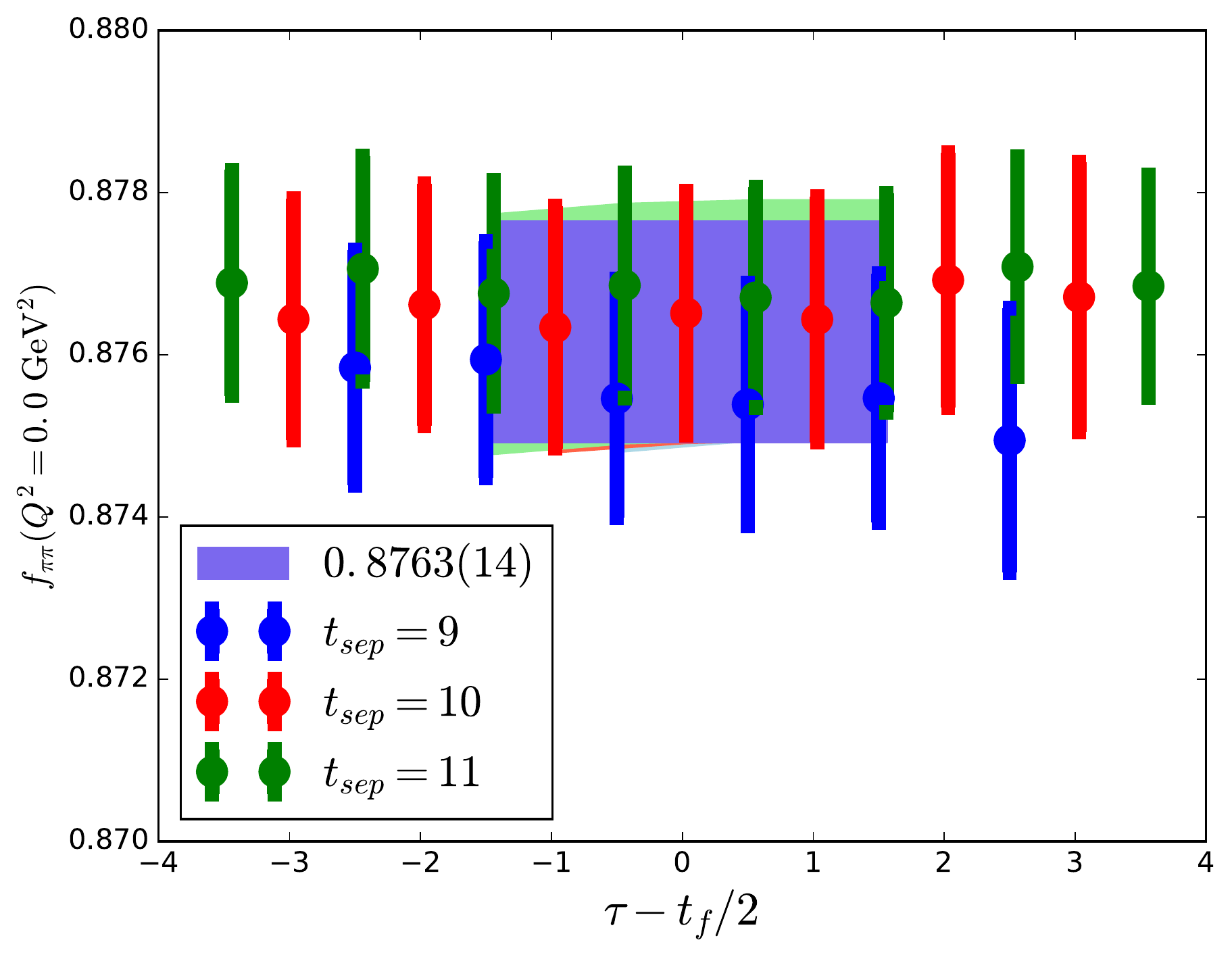}
  \includegraphics[width=0.45\linewidth]{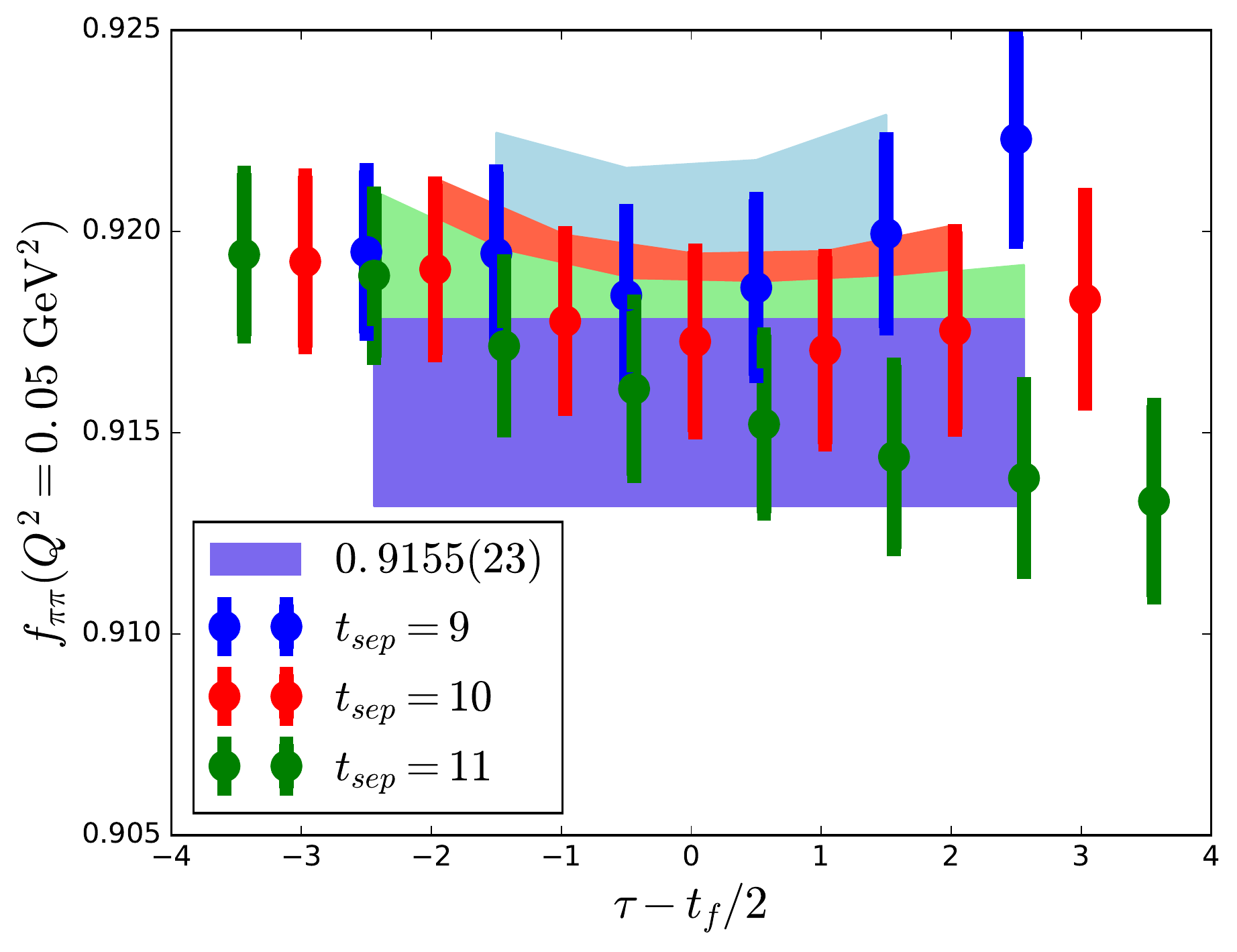}
  \caption{(Left panel) The joint fitting of correlation functions on 32ID at zero momentum transfer. (Right panel) The case of $p_x=0, p_y=0,p_z=2 \pi/L$ with the summation of $8$ symmetric momenta which give momentum transfer $Q^2= \ 0.05 \  {\rm{GeV}}^2$ on the 32ID at pion mass $174$ MeV.}
  \label{fig:32IDV}
\end{figure}

    Precise simulation of the three-point functions with connected-inserstion of vector current $V_4$ for the matrix element $\bra{\pi(p_f)} V_4 \ket{\pi(p_i)}$ has been done with the stochastic sandwich method described in reference~\cite{qcd_collaboration_stochastic_2016}. Along with sink momentum $\vec{p}_f=0$ and mixed momentum source~\cite{qcd_collaboration_stochastic_2016} with momentum $\vec{p}_i$ equal to momentum transfer $\vec{q} = \vec{p}_f - \vec{p}_i$, we can code with the same trick as the sink-sequential method~\cite{bernard_lattice_1985-1,martinelli_lattice_1989} to determine all the momentum transfers with only one three-point function contraction to significantly lower the contraction cost. The corresponding three-point function can be fitted with Eq.~(\ref{eq:3pt})

\begin{eqnarray} \label{eq:3pt}
\begin{aligned}
C_{3pt}(\tau,t_f,\vec{p}_i,\vec{p}_f) = & \sum_{\vec{x}_f,\vec{z}} e^{-i \vec{p}_f \cdot (\vec{x}_f - \vec{z})} e^{-i \vec{p}_i \cdot \vec{z}} \braket{{\rm{T}    }[\chi_{\pi^+}(\vec{x}_f,t_f) V_4(\vec{z},\tau) \chi_{\pi^+}^{\dagger}(\vec{0},[0,T/2])]} \\
\approx & \frac{Z_{\vec{p}_i} Z_{\vec{p}_f} (2 m)^2}{(La)^3 4 E_i E_f} \bra{\pi(p_f)} V_4 \ket{\pi(p_i)} \\
&\times e^{-E_i \tau -E_f(t_f-\tau)} (1 + e^{- E_i T/2}) \\
&+ C_1 e^{-E^1_i \tau - E^1_f (t_f - \tau)} + C_2 e^{-E^1_i \tau -E^1_f (t_f - \tau)} \\
\end{aligned}
\end{eqnarray}
in which common parameters have the same meaning as in the two-point function, $E_i$ and $E_f$ are the pion energy at momentum $\vec{p}_i$ and $\vec{p}_f$ and $E^1_i$ and $E^1_f$ are the corresponding energies of the first excited state. $C_1$ and $C_2$ are the unknown transition amplitudes with corresponding overlap factors from ground state to the first excited state as $\bra{\pi^1(p_f)} V_4 \ket{\pi(p_i)}$ and $\bra{\pi(p_f)} V_4 \ket{\pi^1(p_i)}$, respectively. We ignore the matrix element $\bra{\pi^1(p_f)} V_4 \ket{\pi^1(p_i)}$ which is small in most cases.  By using the same smeared source and smeared sink as in the two-point function, joint fitting of $C_{3pt}$ and $C_{2pt}$ can give the matrix elements $\bra{\pi(p_f)} V_4 \ket{\pi(p_i)}$ at several momentum transfers. In order to control the systematic errors coming from the excited-state contamination, we have done calculations with source-sink separation $t_f-t_i =8a,10a$ and $12a$ on 24I (equal to $0.88\ {\rm{fm}}, 1.10\ {\rm{fm}}$ and $1.32\ {\rm{fm}}$, respectively) and $t_f-t_i = 9a,10a$ and $11a$ on 32ID (equal to $1.287\ {\rm{fm}}, 1.430\ {\rm{fm}}$ and $1.573\ {\rm{fm}}$, respectively). As an example, Fig.~\ref{fig:32IDV} shows the fitting plot on 32ID for pion mass $174 \ {\rm{MeV}}$ at zero momentum transfer and pion form factor $f_{\pi \pi}(Q^2)$ from Eq.~(\ref{eq:fpipi}) at the smallest momentum transfer with $Q^2 \equiv q^2 - (E_f-E_i)^2 = \ 0.05 \  {\rm{GeV}}^2$ with reasonable $\chi^2$ of $0.6-1.2$. The result shows only a $0.5 \%$ difference between separations which means that the current source-sink separations are large enough so that there is a good control of the systematic errors from the excited-state contamination.

\begin{equation} \label{eq:fpipi}
f_{\pi \pi}(Q^2) = \frac{1}{E_i+E_f} \frac{\bra{\pi(p_f)} V_4 \ket{\pi(p_i)}}{\bra{\pi(p_f)} V_4 \ket{\pi(p_f)}}
\end{equation}

    As we are using overlap fermions which have exact chiral symmetry on lattice, we should have vector renormalization constant equal to axial normalization constant, $Z_V = Z_A$. A detailed study of renormalization constants of overlap quark bilinear operators~\cite{qcd_collaboration_ri/mom_2018} shows that $ Z_V/Z_A = 1$ is well satisfied. In reference~\cite{liang_quark_2018}, $Z_A$ has been calculated using the same method in reference~\cite{qcd_collaboration_ri/mom_2018} on the 32ID lattice. In this pion form factor calculation on 32ID, $Z_V$ can be obtained using Eq.~(\ref{eq:ZV}).

\begin{equation} \label{eq:ZV}
Z_V = \frac{1}{\bra{\pi(0)} V_4 \ket{\pi(0)}} \\
\end{equation}

\begin{figure}
  \centering
  \includegraphics[width=0.6\textwidth]{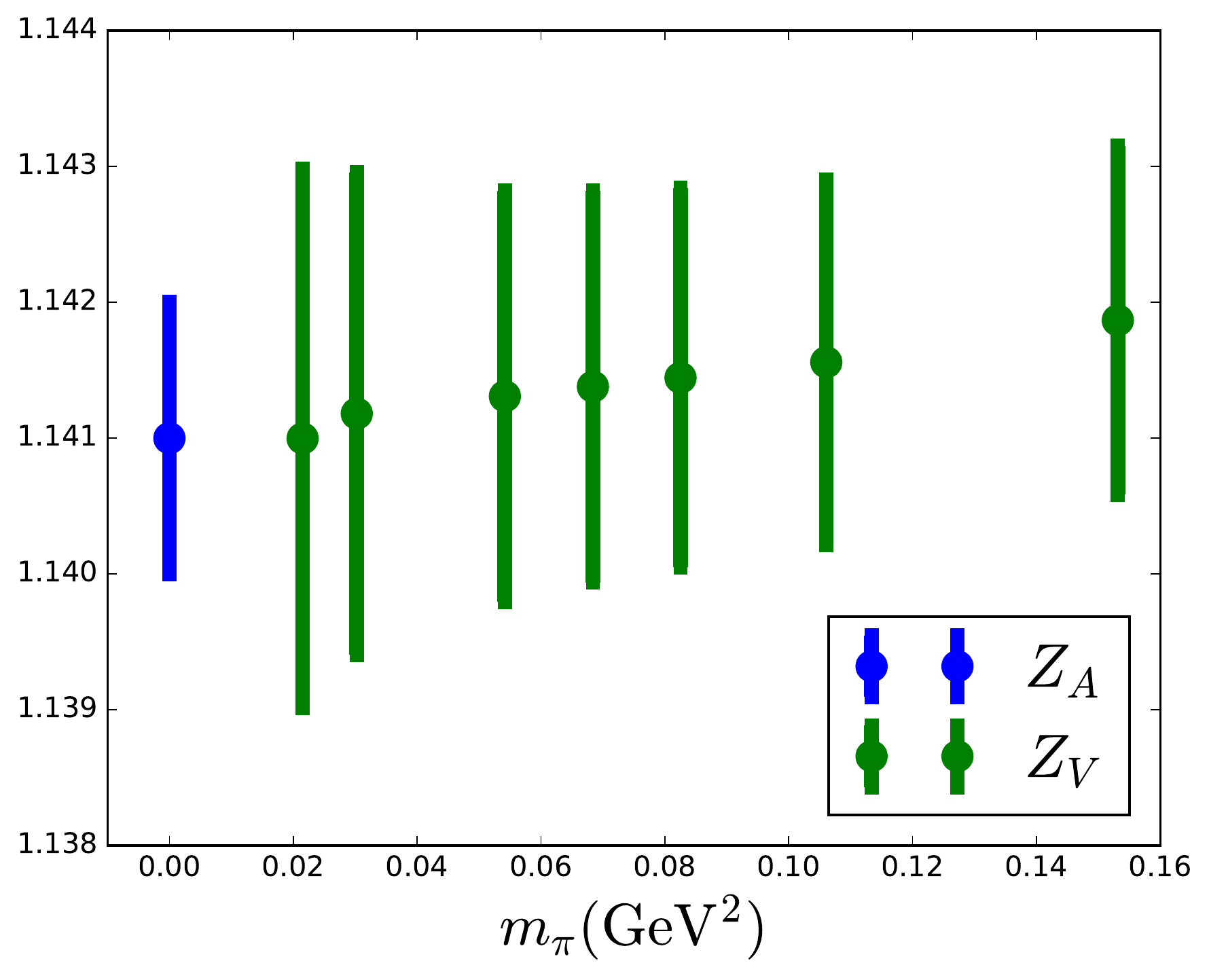}
  \caption{$Z_A$ and $Z_V$ on 32ID}
  \label{fig:ZAV}
\end{figure}

    By comparing the $Z_A$ from ratios of the two-point functions along with the $Z_V$ at seven pion masses from the joint fit of the three-point function and the two-point function in Fig.~\ref{fig:ZAV} on 32ID, it is obvious that within $0.3 \%$ error, $Z_V$ and $Z_A$ agree with each other. With these two different independent analyses, we confirm that current calculation and the joint fitting analysises have good control of various systematic errors.

\begin{figure}
  \centering
  \includegraphics[width=0.6\textwidth]{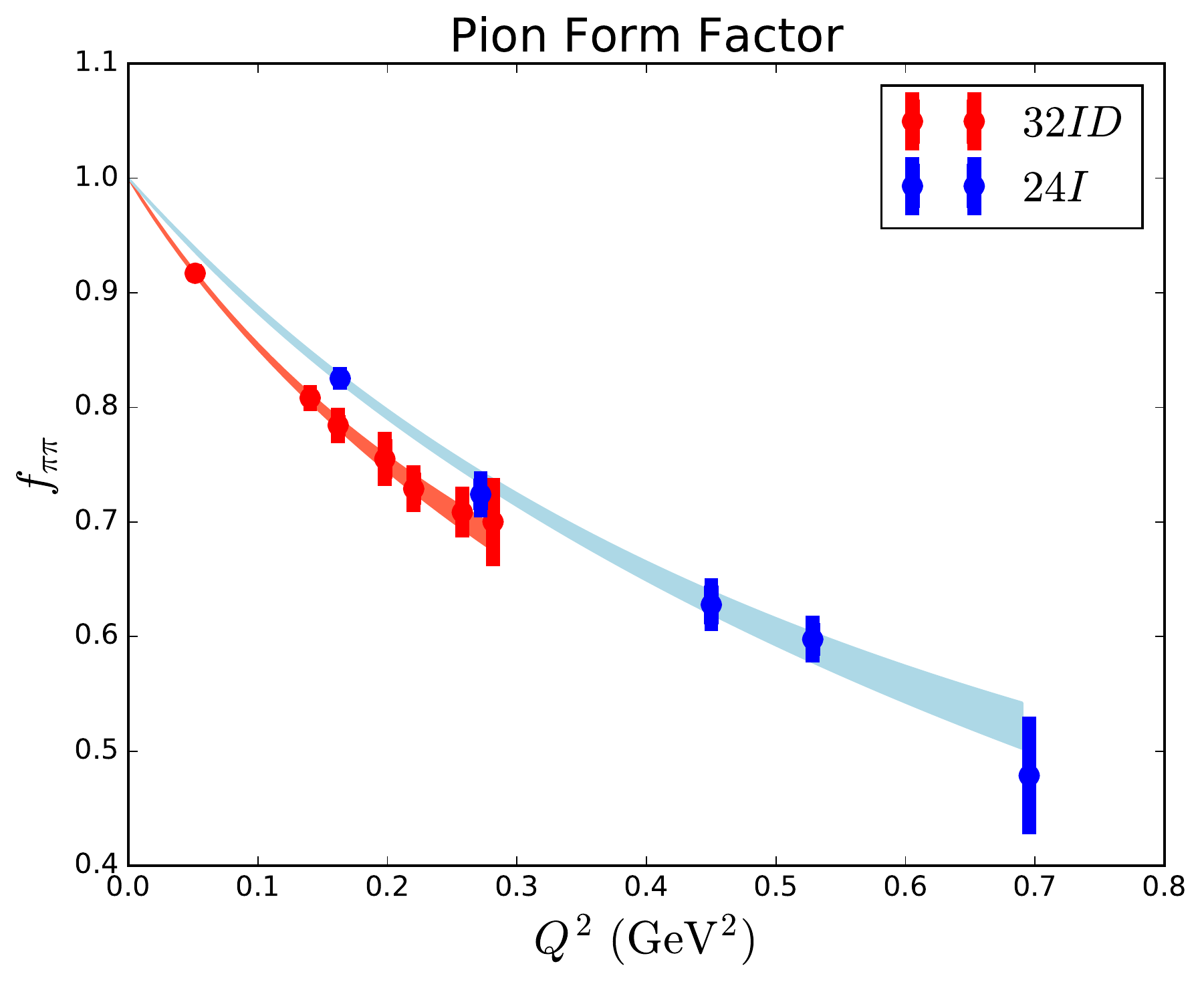}
  \caption{$z$-expansion fitting of the pion form factor on 24I with pion mass $347 \ {\rm{MeV}}$ and on 32ID with pion mass $174 \ {\rm{MeV}}$.}
  \label{fig:z-exp}
\end{figure}

    With the $z$-expansion~\cite{lee_extraction_2015} method we have done a model-independent fitting of the pion form factor $f_{\pi \pi}(Q^2)$ using Eq.~(\ref{eq:z-exp}) with $k_{max} = 2$.
\begin{eqnarray} \label{eq:z-exp}
\begin{aligned}
f_{\pi \pi}(Q^2) &= 1 + \sum_{k=1}^{k_{max}} a_k z^k \\
z(t,t_{cut},t_0) &= \frac{\sqrt{t_{cut}-t_0} - \sqrt{t_{cut}-t_0}}{\sqrt{t_{cut}-t_0} + \sqrt{t_{cut}-t_0}} \\
t = &-Q^2, t_{cut} = 4 m_\pi^2 \\
\end{aligned}
\end{eqnarray}
The pion form factor results are shown in Fig.~\ref{fig:z-exp} with pion mass $347 \ {\rm{MeV}}$ on 24I and $174 \ {\rm{MeV}}$ on 32ID chosen to be close to their domain-wall sea pion mass. We have used $Z_V$ to normalize the form factors at non-zero momentum transfers. With increasing momentum transfers, signals of the matrix elements decrease dramatically so that we can have six or seven momentum transfers which constrain the current fitting parameters. This limits our ability to access higher momenta on coarser and larger lattices such as 32ID compared to 24I.

\begin{table}
  \centering{
  \begin{tabular}{| c | c | c | c | c | c | c | c |}
    \hline
    Lattice & \multicolumn{6}{c}{Pion ({\rm{MeV}})}  & \\
    \hline
    24I  &            &            & 253.26(47) & 281.56(44) & 320.72(43) & 347.38(42) & 388.92(42) \\
    \hline
    32ID & 146.66(19) & 173.95(18) & 232.75(17) & 261.50(17) & 287.28(17) & 325.68(17) & 391.34(17) \\
    \hline
  \end{tabular}
  \caption{ Fitted pion masses on 24I and 32ID }
  \label{tab:pionmass}
  }
\end{table}

    A similar fitting procedure has been done on all valence pion masses listed in Table~\ref{tab:pionmass}. After the $z$-expansion fitting, we can get $\langle r^2 \rangle$ by taking the derivative of the fitted function Eq.~(\ref{eq:radius}).
\begin{equation} \label{eq:radius}
\langle r^2 \rangle \equiv 6 \frac{f_{\pi \pi} (Q^2)}{d Q^2}|_{Q^2 = 0}
\end{equation}
\begin{figure}
  \centering
  \includegraphics[width=0.6\textwidth]{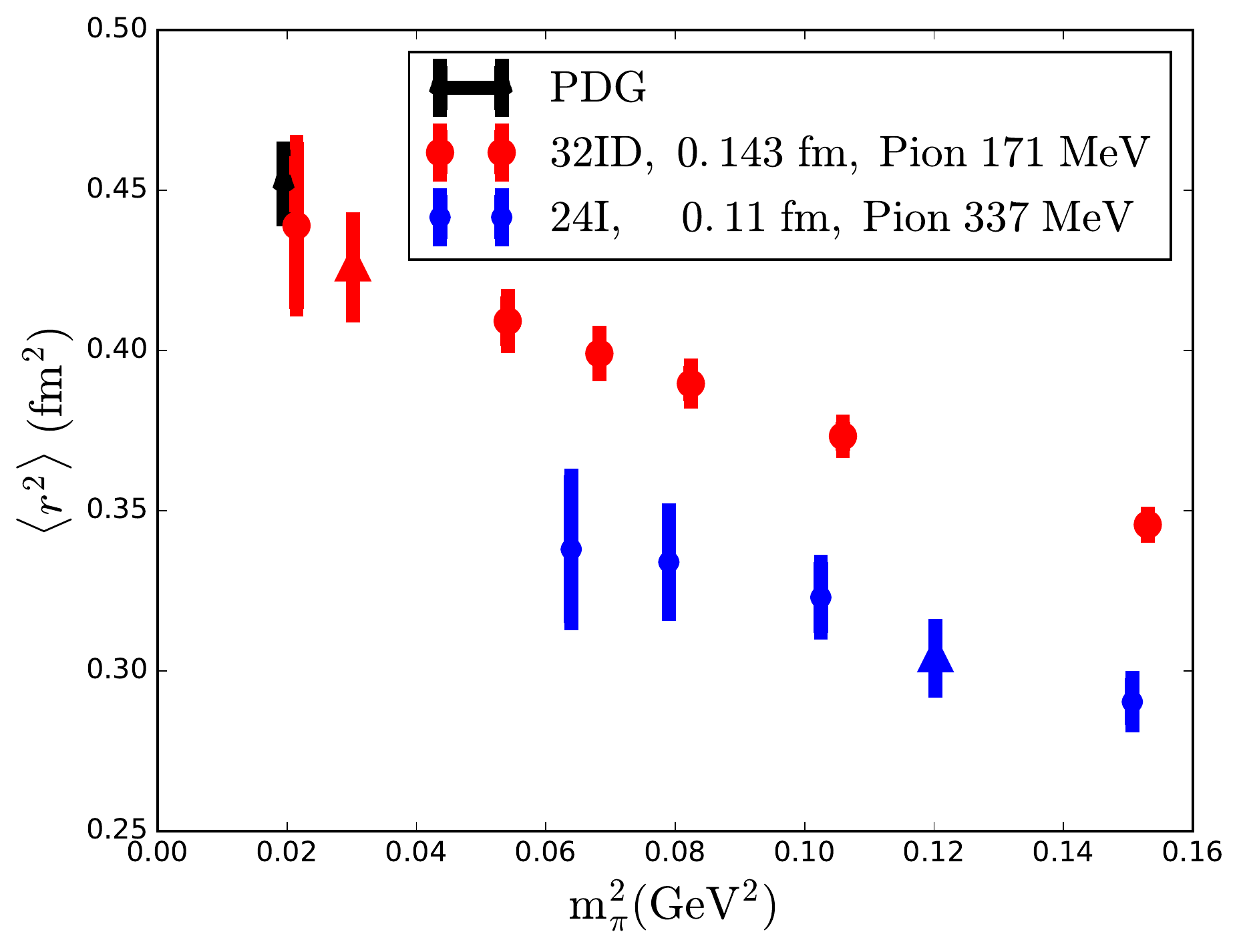}
  \caption{$\langle r^2 \rangle$ as function of $m_\pi^2$ on 24I and 32ID compared to the PDG value}
  \label{fig:global}
\end{figure}
The final $\langle r^2 \rangle$ results on 24I and 32ID are shown in Fig.~\ref{fig:global} with several pion masses including $146.66(19) \ {\rm{MeV}}$ which is very close to physical pion mass $139.57 \ {\rm{MeV}}$. Such lowest pion mass result from 32ID is consistent with the experiment (PDG) value within current statistics. The most notable difference is that the 24I result is around $20 \%$ lower than the 32ID result at similar pion mass. Such a difference may come from their difference of sea pion mass, lattice spacing or lattice volume which is out of control in the current calculation. Due to such a large discrepancy, we will need one or two more comparable lattices to find out which parameter has the largest effect.

\section{Summary and Conclusion}
    We presented calculations of the pion form factor using overlap fermions with a range of pion masses. By differentiating the fitted $f_{\pi \pi}(Q^2)$ shown in Fig.~\ref{fig:z-exp}, we get the corresponding pion charge radius $\langle r^2 \rangle$ to be $0.304(10) \ {\rm{fm^2}}$ on 24I at pion mass $347.38(42) \  {\rm{MeV}}$ and $0.426(15) \ {\rm{fm^2}}$ on 32ID at pion mass $146.66(19) \ {\rm{MeV}}$. With multi-mass algorithm using overlap fermions, we obtained the dependence between pion charge radius and valence pion masses shown in Fig.~\ref{fig:global}. Further calculations on additional lattices with different sea quark masses, lattice spacings and lattice volumes will be needed to do a global fitting to extrapolate to the physical limit.

\bibliographystyle{h-physrev}
%\bibliography{mylibrary}{}
\bibliography{./pioncite.bib}

\begin{thebibliography}{10}

\bibitem{PhysRevD.98.030001}
Particle Data Group, M.~Tanabashi {\em et~al.},
\newblock {\em Review of {{Particle Physics}}},
\newblock Phys. Rev. D {\bf 98}, 030001 (2018).

\bibitem{brommel_pion_2007-1}
D.~Br\"ommel {\em et~al.},
\newblock {\em The Pion Form Factor from Lattice {{QCD}} with Two Dynamical
  Flavours},
\newblock The European Physical Journal C {\bf 51}, 335 (2007),
  hep-lat/0608021.

\bibitem{boyle_pions_2008-1}
P.~A. Boyle {\em et~al.},
\newblock {\em The Pion's Electromagnetic Form Factor at Small Momentum
  Transfer in Full Lattice {{QCD}}},
\newblock Journal of High Energy Physics {\bf 2008}, 112 (2008), 0804.3971.

\bibitem{nguyen_electromagnetic_2011-1}
O.~H. Nguyen, K.-I. Ishikawa, A.~Ukawa, and N.~Ukita,
\newblock {\em Electromagnetic Form Factor of Pion from {{N}}\_f=2+1 Dynamical
  Flavor {{QCD}}},
\newblock Journal of High Energy Physics {\bf 2011} (2011), 1102.3652.

\bibitem{jlqcd_collaboration_pion_2012}
J.~Collaboration {\em et~al.},
\newblock {\em Pion Form Factors in the Epsilon Regime},
\newblock arXiv:1211.0743 [hep-lat]  (2012), 1211.0743.

\bibitem{bali_nucleon_2013-1}
G.~S. Bali {\em et~al.},
\newblock {\em Nucleon Mass and Sigma Term from Lattice {{QCD}} with Two Light
  Fermion Flavors},
\newblock Nuclear Physics B {\bf 866}, 1 (2013), 1206.7034.

\bibitem{brandt_pion_2013-1}
B.~B. Brandt, A.~Juttner, and H.~Wittig,
\newblock {\em The Pion Vector Form Factor from Lattice {{QCD}} and {{NNLO}}
  Chiral Perturbation Theory},
\newblock Journal of High Energy Physics {\bf 2013} (2013), 1306.2916.

\bibitem{jlqcd_collaboration_light_2016}
J.~Collaboration {\em et~al.},
\newblock {\em Light Meson Electromagnetic Form Factors from Three-Flavor
  Lattice {{QCD}} with Exact Chiral Symmetry},
\newblock Physical Review D {\bf 93} (2016), 1510.06470.

\bibitem{koponen_size_2016}
J.~Koponen, F.~Bursa, C.~T.~H. Davies, R.~J. Dowdall, and G.~P. Lepage,
\newblock {\em The Size of the Pion from Full Lattice {{QCD}} with Physical
  \$u\$, \$d\$, \$s\$ and \$c\$ Quarks},
\newblock Physical Review D {\bf 93} (2016), 1511.07382.

\bibitem{brandt_electromagnetic_2013}
B.~B. Brandt,
\newblock {\em The Electromagnetic Form Factor of the Pion: {{Results}} from
  the Lattice},
\newblock International Journal of Modern Physics E {\bf 22}, 1330030 (2013),
  1310.6389.

\bibitem{boyle_low_2016}
P.~A. Boyle {\em et~al.},
\newblock {\em The {{Low Energy Constants}} of \${{SU}}(2)\$ {{Partially
  Quenched Chiral Perturbation Theory}} from \${{N}}\_\{f\}=2+1\$ {{Domain Wall
  QCD}}},
\newblock Physical Review D {\bf 93} (2016), 1511.01950.

\bibitem{li_overlap_2010}
A.~Li {\em et~al.},
\newblock {\em Overlap Valence on 2+ 1 Flavor Domain Wall Fermion
  Configurations with Deflation and Low-Mode Substitution},
\newblock Physical Review D {\bf 82}, 114501 (2010).

\bibitem{degrand_wave-function_1991}
T.~A. DeGrand and R.~D. Loft,
\newblock {\em Wave-Function Tests for Lattice {{QCD}} Spectroscopy},
\newblock Computer Physics Communications {\bf 65}, 84 (1991).

\bibitem{allton_lattice_1991}
C.~R. Allton, C.~T. Sachrajda, V.~Lubicz, L.~Maiani, and G.~Martinelli,
\newblock {\em A Lattice Computation of the Decay Constant of the {{B}}-Meson},
\newblock Nuclear Physics B {\bf 349}, 598 (1991).

\bibitem{qcd_collaboration_stochastic_2016}
{$\chi$QCD Collaboration} {\em et~al.},
\newblock {\em Stochastic Method with Low Mode Substitution for Nucleon
  Isovector Matrix Elements},
\newblock Physical Review D {\bf 93}, 034503 (2016).

\bibitem{bernard_lattice_1985-1}
C.~Bernard, T.~Draper, G.~Hockney, A.~M. Rushton, and A.~Soni,
\newblock {\em Lattice {{Calculation}} of {{Weak Matrix Elements}}},
\newblock Physical Review Letters {\bf 55}, 2770 (1985).

\bibitem{martinelli_lattice_1989}
G.~Martinelli and C.~T. Sachrajda,
\newblock {\em A Lattice Study of Nucleon Structure},
\newblock Nuclear Physics B {\bf 316}, 355 (1989).

\bibitem{qcd_collaboration_ri/mom_2018}
{$\chi$QCD Collaboration} {\em et~al.},
\newblock {\em {{RI}}/{{MOM}} and {{RI}}/{{SMOM}} Renormalization of Overlap
  Quark Bilinears on Domain Wall Fermion Configurations},
\newblock Physical Review D {\bf 97}, 094501 (2018).

\bibitem{liang_quark_2018}
J.~Liang, Y.-B. Yang, T.~Draper, M.~Gong, and K.-F. Liu,
\newblock {\em Quark Spins and {{Anomalous Ward Identity}}},
\newblock arXiv:1806.08366 [hep-lat, physics:hep-ph]  (2018), 1806.08366.

\bibitem{lee_extraction_2015}
G.~Lee, J.~R. Arrington, and R.~J. Hill,
\newblock {\em Extraction of the Proton Radius from Electron-Proton Scattering
  Data},
\newblock Physical Review D {\bf 92} (2015), 1505.01489.

\end{thebibliography}
%\bibliography{/home/ard/Back_Up/Lattice_data/Bibtex/Pos.bib}

\end{document}